\documentclass[aip,apl,graphicx]{revtex4-1}

\draft 

\usepackage[utf8]{inputenc}
\usepackage[T1]{fontenc}
\usepackage{mathptmx}
\usepackage{etoolbox}
\usepackage{graphicx}
\usepackage{gensymb}
\usepackage{amsmath}
\usepackage{lipsum}
\usepackage[dvipsnames]{xcolor}
\usepackage{soul}

\graphicspath{ {./images/} }

\def\dI{\textrm{d}I}
\def\dV{\textrm{d}V}
\makeatletter
\def\@email#1#2{
 \endgroup
 \patchcmd{\titleblock@produce}
  {\frontmatter@RRAPformat}
  {\frontmatter@RRAPformat{\produce@RRAP{*#1\href{mailto:#2}{#2}}}\frontmatter@RRAPformat}
  {}{}
}
\makeatother

\begin{document}


\title{Magnetic molecule tunnel heterojunctions} 



\author{Xuanyuan Jiang}
\affiliation{Department of Physics, University of Florida, Gainesville Fl}

\author{Andrew Brooks}
\affiliation{Department of Physics, University of Florida, Gainesville Fl}

\author{Shuanglong Liu}
\affiliation{Department of Physics, University of Florida, Gainesville Fl}

\author{John Koptur-Palenchar}
\affiliation{Department of Physics, University of Florida, Gainesville Fl}

\author{Yundi Quan}
\affiliation{Department of Materials Science and Engineering, University of Florida, Gainesville Fl}

\author{Richard G. Hennig}
\affiliation{Department of Materials Science and Engineering, University of Florida, Gainesville Fl}

\author{Hai-Ping Cheng}
\affiliation{Department of Physics, University of Florida, Gainesville Fl}

\author{Xiaoguang Zhang}
\affiliation{Department of Physics, University of Florida, Gainesville Fl}

\author{Arthur F. Hebard*}
\email[]{afh@phys.ufl.edu}
\affiliation{Department of Physics, University of Florida, Gainesville Fl}


\date{\today}

\begin{abstract}
We characterize molecular magnet heterojunctions in which sublimated CoPc films as thin as 5 nm are sandwiched between transparent conducting bottom-layer indium tin oxide and top-layer soft-landing eutectic GaIn (EGaIn) electrodes. The roughness of the cobalt phthalocyanine (CoPc) films was determined by atomic force microscopy to be on the order of several nanometers, and crystalline ordering of lying-down planar
molecules was confirmed by X-ray diffraction. The current-voltage (I-V) characteristics reveal the onset of a superconducting gap at $T_c =6$\thinspace K, which together with higher temperature fits to a modified Simmons' model, provide incontrovertible evidence for direct quantum mechanical tunneling processes through the magnetic molecules in our heterojunctions. The voltage dependent features in the differential conductance measurements relate to spin states of single molecules or aggregates of molecules and should prove to be important for quantum information device development.
\end{abstract}


\maketitle


Molecular magnets (MMs) are promising as the basic building blocks of quantum information systems. With chemically designed ligand surroundings, MMs can be identically replicated, assembled into linked nanoscale aggregates either in bulk or in thin films, tuned with external fields, and found to manifest quantum properties such as superimposition of quantum states and the preservation of quantum coherence.\cite{Hill_1,Atzori_JACS,Wasielewski,Christou_Polyhedron} The nascent field of molecular spintronics merges the advantages of spintronics where electrical currents comprising spin-up and spin-down carriers interact with MMs\cite{Bogani_Wernsdorfer,MPc_spintronic} which by themselves can act as robust spin qubits for storage and processing of quantum information.\cite{Ding} Many of the pioneering studies of MMs using techniques such as electron paramagnetic resonance, inelastic neutron scattering and quantum tunneling have been restricted to bulk samples in the form of crystals, powders or frozen solutions.\cite{Hill_Dalton,Aravena_Dalton} Studies of single MMs absorbed on surfaces have revealed a wealth of information relevant to QIS systems but are limited to using scanning probe microscope tips as electrodes\cite{TbPc2,MMs_graphene} or to nanoscale break junctions coupled to source, drain and gate electrodes.\cite{nuc1} Incorporating MMs into practical  device applications where multiple MMs must be addressed requires eliminating the bulky STM tip as an electrode and sandwiching the MMs placed either as single entities or as arrays between two thin-film electrodes. 

Here we report on our progress in fabricating prototypes of magnetic molecule tunnel heterojunctions using soft landing counter electrodes which can be cycled to temperatures as low as 2\thinspace K without loss of electrical contact. By eliminating the STM tip as an electrode, our work shows promise for the incorporation of fragile MMs into practical thin-film device applications. We have chosen as our magnetic molecules commercially available metallophthalocyanine ions - in particular CoPc, which can be easily sublimated without disassociation into ultrathin films that have well studied magnetic properties.\cite{CoPc_chains,CoPc_highTAFchains,MPc_spintronic} A critical challenge is in the choice of a soft-landing counterelectrode which does not damage the fragile MMs and cause pinhole shorts as does for example the deposition of thermally sublimated metals onto CoPc films less than 15 nm thick.\cite{MPc_microshorts} We follow the example of investigators who use liquid eutectic soft landing GaIn (EGaIn) counter electrodes\cite{Chiechi_EGaIn,Mol_ensembles} to study charge transport through molecular ensemble junctions sandwiched between conducting electrodes. The types of molecules studied include aliphatic and aromatic hydrocarbons, biomolecular ensembles, and long chain self-assembled monolayers (SAMs) with various substrate anchoring groups. The consensus of these works is that  tunneling as described by the Simmons model\cite{Simmons1,Simmons2} and thermally activated hopping processes both contribute to charge transport.\cite{Mol_ensembles}

\begin{figure}[b]
\includegraphics[width=1\linewidth]{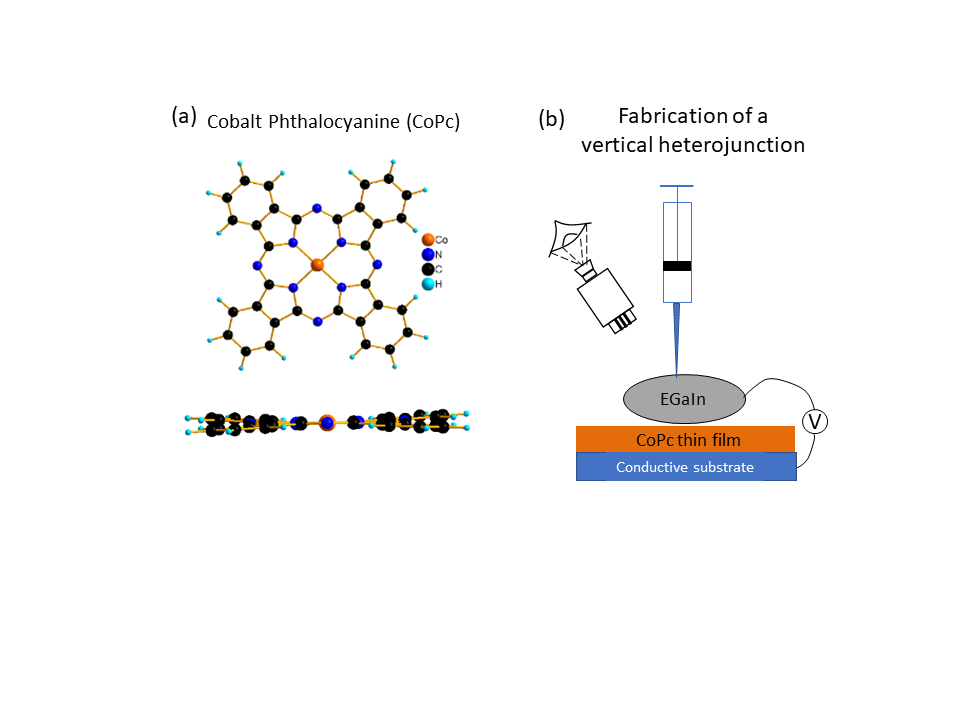}
\centering
\caption{(a) Molecular structure of cobalt phthalocyanine (CoPc). (b) Schematic diagram of the CoPc vertical heterojunction fabrication.}
\label{fig:sublimation}
\end{figure}

Variation of sublimation rates and substrate temperatures during deposition allow us to obtain MM films, which as verified by AFM and X-ray diffraction have 100\% areal coverage, thicknesses $d \leq 5$\thinspace nm and crystalline structure. In choosing CoPc molecules, we benefit from studies using AFM/STM and SQUID magnetometry which reveal magnetic properties of CoPc MMs absorbed on a variety of surfaces and in powder form.\cite{CoPc_chains,MPc_spintronic,CoPc_highTAFchains}  
Using current-voltage (I-V) and differential conductance ($\dI/\dV$-$V$) measurements, we confirm quantum mechanical tunneling from high temperature (40-200\thinspace K) Simmons model fits and a low temperature ($T_c = 6$\thinspace K) opening of a superconducting gap associated with the metastable $\beta$ phase of Ga in the EGaIn contact.\cite{Feder,Hirshfeld_Hennig}. These manifestations of quantum mechanical tunneling in our ultra-thin heterojunctions together with our observation of $\dI/\dV$ peaks associated with the MM layer residing between the electrodes promises an all thin film technology for QIS devices utilizing MMs without the need for impractical and expensive STM probes as electrodes. 

Fabrication steps in the formation of our heterojunctions begin as shown in the schematic of Fig. \ref{fig:sublimation} with the sublimation of CoPc onto smooth conducting substrates: Pt with root-mean-square roughness less than 1\thinspace nm or ITO  with roughness less than 3\thinspace nm. The CoPc (Sigma-Aldrich) is evaporated at a sublimation temperature near 600\thinspace K at rates on the order of 1\thinspace \text{\normalfont\AA} per minute onto the bottom electrode at a temperature which can be regulated to $\pm 1$\thinspace K over a temperature range from 77\thinspace K to 400\thinspace K. For 5\thinspace nm thick CoPc films, the films become progressively smoother with substrate temperature increasing from 100\thinspace K (discontinuous films) to 325\thinspace K (smooth films with root-mean-square roughness of 2.9\thinspace nm). These data are shown in Fig. S1 of Supplemental Information. 

\begin{figure}[b]
\includegraphics[width=1.0\linewidth]{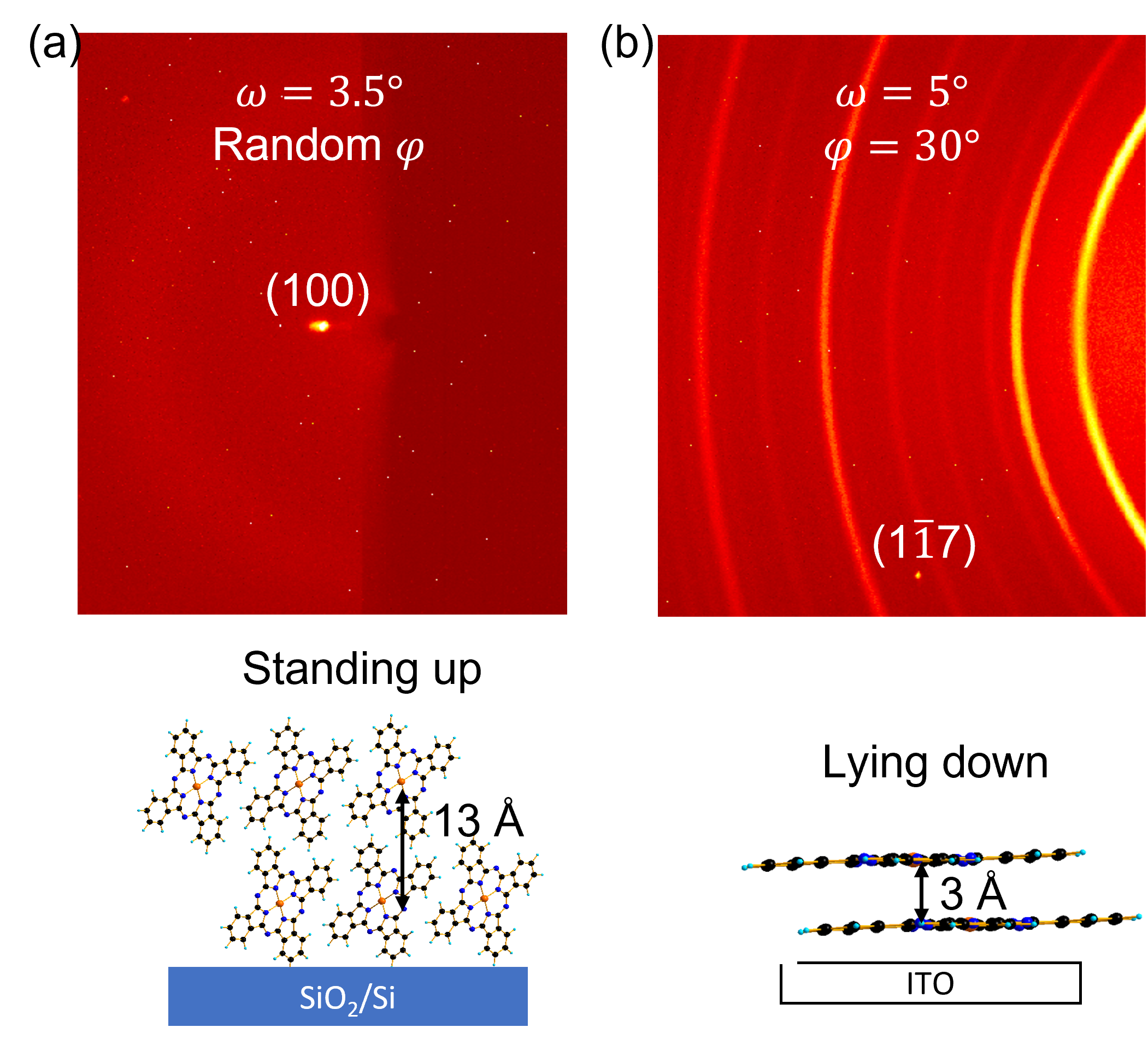}
\centering
\caption{XRD images of 5 nm CoPc grown on different substrates. (a) A 2D XRD pattern of 5 nm CoPc on SiO2/Si. The crystal plane (100) is captured when the incident beam is at 3.5\degree and the sample rotation angle is random. (b) A 2D XRD pattern of 5 nm CoPc on ITO. The crystal plane (1$\bar{1}$7) is captured when the incident beam is at 5\degree \thinspace and the sample rotation angle is at 30\degree.}
\label{fig:XRD}
\end{figure}

X-ray study of our deposited films is motivated by the recognition that if the MM thin film is in amorphous or polycrystalline form, the magnetic moments of randomly oriented molecules (or equivalently, randomly oriented grains in itinerant magnets\cite{Hysteresis_single_vs_polyXtal}) would cancel each other and mask the magnetic properties of isolated single MMs. The crystalline structure of our sublimated films, shown in the diffraction scans of Fig.~\ref{fig:XRD}, is determined using a single crystal X-ray diffractometer in which $\omega$ and $\phi$ scans are independently captured on a 2D detector. Our results can be compared with previous studies\cite{MPc_microshorts,MPc_XRD_Schuller,MPc_orientation,organic_XRD_Schuller} in which a strong influence of the substrate on the orientation of the CoPc molecules is found. The diffraction spots on the 2D detector, indicate that our CoPc thin films are in single crystalline form. However, positions of these diffraction spots are different on different substrates. Our conclusions from the very visible diffraction spots shown in Fig.~\ref{fig:XRD} and discussed further in Figs. S2 and S3 of Supplemental Information are that the CoPc deposited on insulating SiO$_2$/Si (conducting ITO) are standing up (lying down) with a lattice spacing of 13\thinspace \text{\normalfont\AA} (3\thinspace \text{\normalfont\AA}).

The liquid EGaIn electrode is applied at room temperature using a syringe as shown schematically in Fig.~\ref{fig:sublimation}b. An optical microscope focused on the EGaIn droplet facilitates the placement of 15\thinspace $\mu$m diameter Cu contact wires within each drop which typically have disk shaped contact areas on the underlying substrate with diameters in the range 200 to 400\thinspace $\mu$m. The macroscopic size of the contact area compared to the microscopic size of a planar CoPc molecule (1.7$\times 10^{-14}$\thinspace cm$^2$) implies that all measurements for these large area junctions\cite{Mol_Elect} represent ensemble averages\cite{Mol_ensembles} with meaningful results obtained only when the molecules between the electrodes are aligned and thickness variations together with the presence of defects is minimized. Spin crossover transitions in ensembles of magnetic molecules embedded in EGaIn heterojunctions have been detected by transport current measurements in 6.7\thinspace nm thick films near room temperature\cite{VTJ_SCO} and by capacitance measurements near 120\thinspace K in 20\thinspace nm thick films.\cite{Shatruk}  

\begin{figure}[b]
\includegraphics[width=1.0\linewidth]{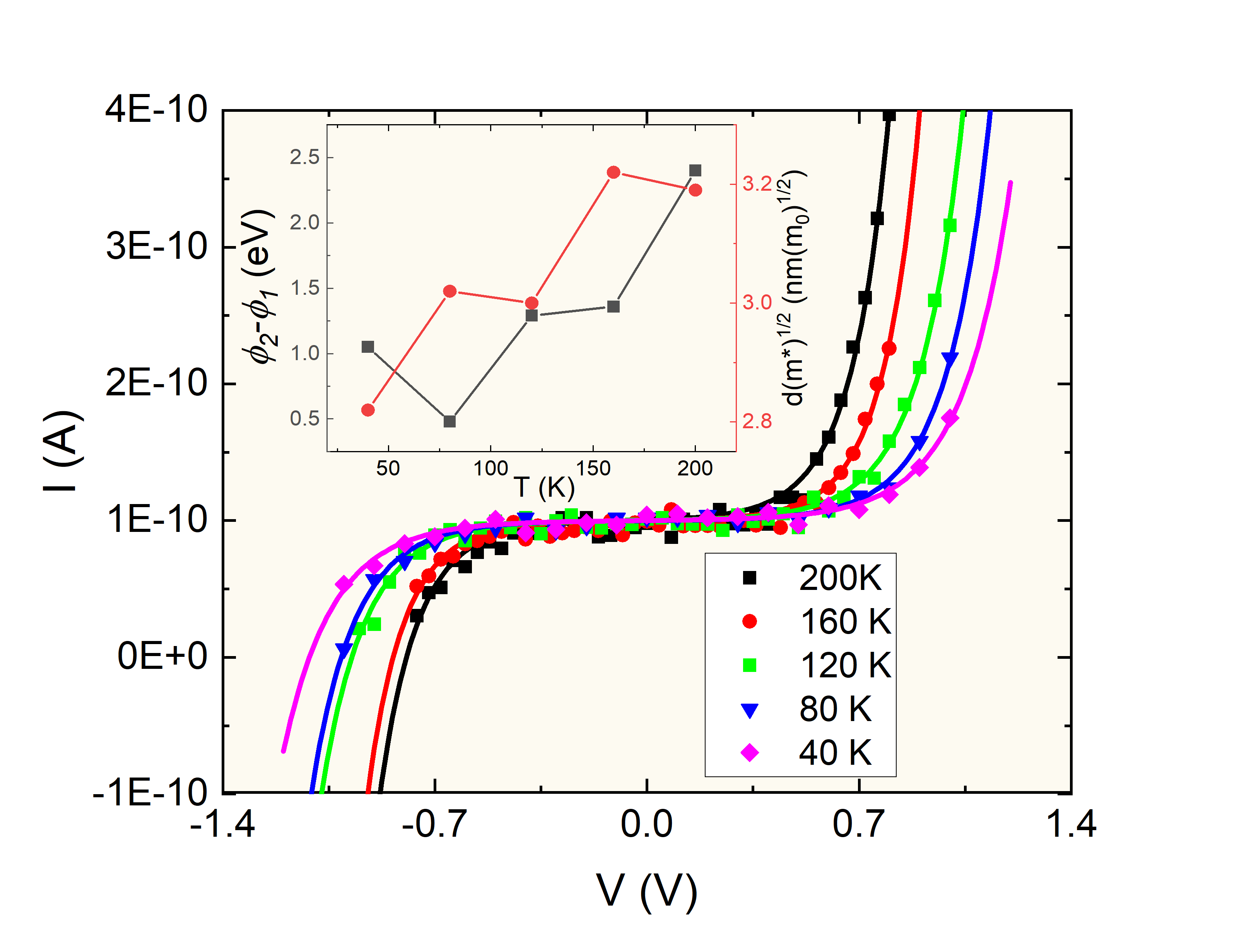}
\centering
\caption{Theoretical fits (solid lines) using the Simmons model discussed in the text describe well the experimental data (symbols) of G$\Omega$ resistance junctions (at $V=0$) measured at the temperatures shown in the legend. The variation with temperature of the fitting parameters $\phi_2-\phi_1$ and $d(m^*)^{1/2}$ over the temperature range 40-200 K is shown in the inset.}
\label{fig:IVs_at_T}
\end{figure}

The initial report\cite{Liquid_metals} that EGaIn electrodes with low contact resistance reliably conform to underlying fragile molecular films without causing damage has led to a flurry of activity in the study of self assembled monolayers (SAMs),\cite{Chiechi_EGaIn} molecular ensembles\cite{Mol_Elect} and advances in microelectronic interconnect technology.\cite{EGaIn_microelectronic} Despite the advantages of low melting temperature, non-toxicity, and the ability to wet most materials without fluxing, \cite{Galinstan,EGaIn_microelectronic} there are problems with corrosion\cite{EGaIn_corrosion} when Pd, Au or Al contact wires are used and reproducibility of tip contact morphology due to formation of Ga$_2$O$_3$ at the surface of the EGaIn during application of the contact.\cite{EGaIn_polished} We have found however that 15\thinspace $\mu$m diameter Cu wires which are completely immersed in the EGaIn contact during application give reliable contacts to temperatures as low as 2\thinspace K, significantly lower than the 110\thinspace K temperatures reported for EGaIn heterojunction contacts by previous workers.\cite{Nijhuis}

The current-voltage (I-V) isotherms for an EGaIn/CoPc(5nm)/ITO heterojunction shown in Fig.~\ref{fig:IVs_at_T} are found to be well described by a generalization\cite{Li_Ma_Wang} of the Simmons’ formula\cite{Simmons1,Simmons2} for an asymmetric barrier tunnel junction, namely:
%
%
\begin{eqnarray}
       J(V) &=& \frac{e}{2\pi hd^2}\left( 1+\frac{(\phi_2-\phi_1-eV)^2}{48\bar{\phi}^2} \right)\left\{
       \left( \phi_-+ \frac{3\sqrt{\phi_-}}{S^{'} d} + \frac{3}{(S^{'} d)^2} \right) 
       \exp{\left[ -S^{'} d\sqrt{\phi_-}\right]}\right. 
       \nonumber \\
    & -& \left.\left( \phi_++ \frac{3\sqrt{\phi_+}}{S^{'} d} + \frac{3}{(S^{'} d)^2} \right)
        \exp{\left[ -S^{'} d\sqrt{\phi_+}\right]}
        \right\}
\end{eqnarray}
%
%
with $ \phi_\pm = \bar{\phi} \pm eV/2$,  $S^{'}=S/(1+(\phi_2-\phi_1-eV)^2/48\bar{\phi}^2)$, $S = 4\pi\sqrt{2m^*}/h$, and $\bar{\phi}$ as the average barrier height. The asymmetric barrier heights may arise due to a dipole layer in the interface \cite{Ishii1999} or, in the absence of charge transfer, to the ‘pillow-effect’. \cite{Ishii1998,Kahn2003} The theory includes higher order terms that are neglected in the original Simmons’ formula but become important for barrier widths less than 10 nm. We interpret the two barrier height parameters $\phi_1 < \phi_2$ as the interface barrier ($\phi_2$) and the effective barrier of the molecular layer ($\phi_1$), respectively. The electron's effective mass, $m^*$, and the barrier width, $d$, are consolidated into a single fitting parameter $d\sqrt{m^*}$. The barrier width in the prefactor is subsumed into the area which multiplies the current density to get the total current. 
%
%
The resulting fits at the indicated temperatures in the legend of Fig.~\ref{fig:IVs_at_T} are shown as solid lines and the variation with temperature of these parameters is shown in the inset. 
%
%
The conversion of the experimentally measured current $I$ to current density $J$ required by theory (Eq. 1) depends on the measured contact area of the junction. 
%
%
Using the 400 $\mu$m diameter of the EGaIn contact for this junction and the measured thickness of 5\thinspace nm, we note that the fitted effective mass falls into the range of $[0.32:\, 0.41]\, m_0$ which is close to the effective mass of ITO measured by optical methods.~\cite{m_ITO} This result implies that most of the junction area is participating in the transport.

\begin{figure}[b]
\includegraphics[width=1.0\linewidth]{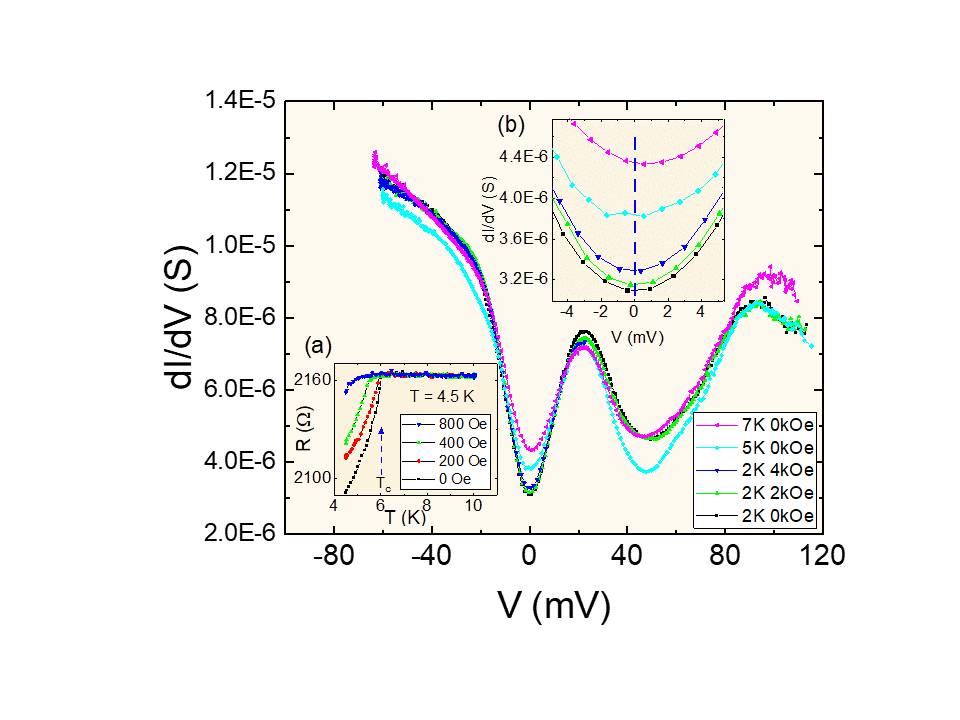}
\centering
\caption{Differential conductance plotted versus dc bias voltage at the indicated temperatures and selected magnetic fields. The EGaIn electrode is biased negatively and voltage modulations are typically on the order of mV. The magnetic field induced suppression of the superconducting transition of the EGaIn electrode is shown in panel (a) for a low resistance (2120 $\Omega$) junction at $T=4.5$\thinspace K. The opening of the superconducting gap as measured by $\dI/\dV$ for the high resistance (320 k$\Omega$) junction is shown in the magnified region (panel (b)) near zero bias.}
\label{fig:dIdV}
\end{figure}

We note that there are no temperature dependencies in the Simmons formula except for those that are extrinsic and might appear in the tunnel barrier width $d(T)$, the effective mass of the electron $m^*(T)$, and the barrier heights $\phi_1(T)$ and $\phi_2(T)$. Accordingly, it is somewhat surprising that the quantum mechanical model described by Eq. 1 describes the high temperature data in Fig.~\ref{fig:IVs_at_T} so well without the inclusion of thermally assisted sequential tunneling  contributions reported for CoPc junctions with thicknesses greater than 15-20\thinspace nm.\cite{MPc_microshorts} We gain additional insight into this behavior by going to lower temperatures and lower junction resistances such as shown in the temperature dependent resistance ($R$-$T$) and differential conductance ($\dI/\dV$-$V$) curves of Fig.~\ref{fig:dIdV}. To obtain the $\dI/\dV$ data, a current ramp with a sinusoidal modulation at 17 Hz in series with a ballast resistor is imposed on the junction with the current $\dI$ and voltage $\dV$ modulations measured by two synchronously coupled lock-in amplifiers. The dc voltage is measured at the output of a low-pass filter with 0.1 Hz roll-off.

The two-terminal resistance transitions of the low resistance (2160 $\Omega$) junction in panel (A) of Fig.~\ref{fig:dIdV} include the series resistance of the EGaIn contact which undergoes a superconducting transition at $T_c=6$\thinspace K. The superconducting $\beta$-Ga phase is a metastable polymorph of Ga.\cite{Bosio:a09881} Gallium can solidify into metastable $\beta$-Ga when the liquid is confined to a micrometer or sub-micrometer scale,\cite{DiCicco} and $\beta$-Ga also forms upon annealing of amorphous Ga at low temperatures.\cite{Bosio:a09881} A recent theoretical study showed that the formation of $\beta$-Ga is kinetically favored over the thermodynamically stable $\alpha$-Ga phase above 174\thinspace K.\cite{Niu} Therefore, the solidification of liquid Ga-In in a eutectic transformation leading to the formation of a lamellar Ga/In microstructure could favor the formation of $\beta$-Ga due to the size confinement. The $\beta$-Ga phase is superconducting with a critical temperature of $T_c$ = 6\thinspace K.\cite{Feder,Hirshfeld_Hennig} In contrast, $\alpha$-Ga and In display much smaller critical temperatures of 0.9 and 3.4\thinspace K, respectively.\cite{Feder,Merriam} The observed superconducting transformation of the In/Ga eutectic at $T_c$ = 6\thinspace K indicates that the confinement in the microstructure during solidification leads to the formation of $\beta$-Ga.

Superconductivity has a more profound effect on the $\dI/\dV$ curves of the 320\thinspace k$\Omega$ junction (Fig.~\ref{fig:dIdV} main and panel (b)) where the series-connected contribution from the much smaller resistance of the EGaIn contact can be ignored. Here the density of states is proportional to $\dI/\dV$ which decreases in the low voltage region as the superconducting gap opens up when $T$ drops below $T_c = 6$\thinspace K. At the lowest available temperature of 2\thinspace K, the sequential application of 2\thinspace kOe and 4\thinspace kOe fields progressively attenuates the gap as would be expected for superconductivty where the gap goes to zero for $T \geq T_c$ and/or magnetic fields greater than the critical field. In addition to magnifying the effect of magnetic field on the zero bias conductance, panel (b) shows the higher temperature shift in symmetry in agreement with the pronounced asymmetry seen at higher voltages in the main panel.

Our observation of the opening of a superconducting gap distinguishes normal metal ohmic shorts, which at the lowest temperatures would not show a superconducting gap, from quantum mechanical tunneling where the presence of a superconducting gap straddling the Fermi energy creates a region where single particle excitations (tunnel electrons) are not allowed. In quantum applications of our heterojunctions, superconductivity in the electrodes could provide an advantage by providing protection of excited spin states \cite{SCprotection_spin_states} and by extending quantum coherence times.\cite{SMM_on_Pb}

\begin{figure}[b]
\includegraphics[width=0.9\linewidth]{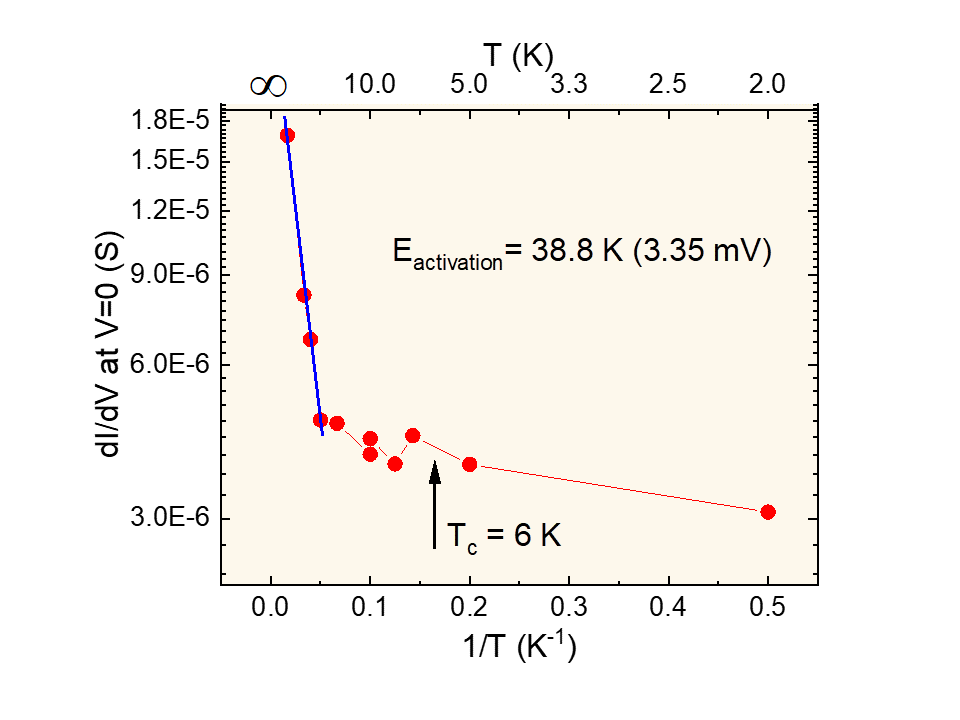}
\centering
\caption{2 Differential conductance $\dI/\dV$ for the same junction in Fig.~\ref{fig:dIdV} evaluated at $V=0$ and plotted on logarithmic axes versus reciprocal temperature shows a crossover from activated behavior (blue line fit) to lower temperatures that include the onset of superconductivity at $T_c = 6$\thinspace K (vertical arrow).}
\label{fig:activation}
\end{figure}

Our detection of superconducting gaps in the I-V characteristics together with the Simmons model fits convincingly establishes quantum mechanical tunneling as the dominant mode of transport in our magnetic molecule heterojunctions.  The explanation for why the high temperature fits in Fig.~\ref{fig:IVs_at_T} work so well for a tunneling model in the absence of competing thermally activated processes is found in the data of Fig.~\ref{fig:activation}. Here the reciprocal temperature dependence of the differential conductance evaluated at zero bias shows activated behavior down to approximately 20 K but with an activation energy, 39K, sufficiently below the temperatures in Fig. 1 to render activated processes unobservable. Said in another way, processes with activation energies lower than the temperature of measurement are difficult to observe. Similar low activation energies on the order of 5-10 meV have been reported for SAM-based junctions with EGaIn top electrodes.\cite{Nijhuis} We note the crossover temperature near 20\thinspace K where weakly activated behavior with $E_{activation} = 3.35$\thinspace K gives way to temperature independent tunneling and associated quantum processes.

The early work on self assembled monolayer (SAM) heterojunctions using EGaIn electrodes
\cite{Chiechi_EGaIn,Mol_ensembles} concludes \cite{Nijhuis,Reus} that for temperatures in the range
$110<T<300$\thinspace K electron transport is due to tunneling as described by the simplified Simmons model.
By comparing results from junctions fabricated from a large variety of SAMs and base electrode choices, these authors conclude that the tunnel transmission probability is dominated by the molecules within the junction rather than the Ga$_2$O$_3$ oxide formed on the surface of the EGaIn or the work functions of the metals.\cite{Nijhuis,Reus}
Our ability to make measurement down to temperatures as low as 2\thinspace K augments these conclusions by revealing a proliferation of molecular signatures, an example of which is shown for CoPc in the main panel of Fig.~\ref{fig:dIdV}. 

The precise identification of the peaks near 20 and 100\thinspace mV in Fig.~\ref{fig:dIdV} is beyond the scope of this study, but there is little doubt that they are related to the molecules within the junction. This particular junction had stable characteristics for one month of measurement although other junctions fabricated on the same thickness MM film showed different zero bias resistances and degrees of asymmetry (or rectification\cite{Reus}) measured by the ratio of voltage dependent forward bias to reverse bias currents. These asymmetries have been attributed\cite{Reus} to molecular orbitals located asymmetrically within the MM layer. Our preliminary measurements on MnPc heterojuntions (see Fig.~S4 of Supplemental Information) show a different set of similar sized peaks. Artifacts due to the electrodes do not seem to be at play since voltage dependent $\dI/\dV$ and $\textrm{d}^2I/\dV^2$ scans on EGaIn/H$_2$Pc(5nm)/ITO shown in Fig.~S5 of Supplemental Information show a proliferation of stable peaks with less asymmetry over a wider range of bias voltages. The H$_2$Pc sample does not contain magnetic transition metal cations.   

The work reported here offers an opportunity to develop magnetic molecule based quantum information systems and thin-film heterostructures while avoiding the use of expensive and impractical scanning probe tips for active \textit{in situ} detection. In providing a road map for further development of applications involving EGaIn based magnetic molecule tunnel heterojunctions, there are some daunting challenges: the optimal molecule or molecules in isolated, dimer, or chain form must be chosen, ultra thin crystalline layers must be engineered, and defects must be minimized to ensure robust reproducible behavior. To obtain in-plane lateral resolution, nanopatterning techniques\cite{Garcia,Hu} using AFM tips to sculpt the bottom electrodes can provide extended scalable structures on $\approx$10 nm length scales. In addition to transparent conducting ITO bottom electrodes, which also allow optical excitation/detection of the MMs, we have had equal success with smooth Pt films. Another possibility is to replace the bottom electrode with a ferromagnetic electrode and use inelastic tunneling spectroscopy ($\dI/\dV$) for the detection of spin states in molecular magnets interacting with spin polarized tunneling currents.\cite{TbPc2} 




%
%

%
\begin{acknowledgments}
 XRD assistance from Dr. Taehoon Kim and Dr. Khalil Abboud in the Center for X-Ray Crystallography and Dr. Kristy Schepker in the Research Service Centers in the University of Florida is gratefully acknowledged. This work is supported as part of the Center for Molecular Magnetic Quantum Materials, an Energy Frontier Research Center funded by the U.S. Department of Energy, Office of Science, Basic Energy Sciences under Award No. DE-SC0019330. 
\end{acknowledgments}

\bibliographystyle{apsrev4-1}
\bibliography{CoPc.bib}

\end{document}